\documentclass{article}
\usepackage[german, english]{babel}
\usepackage[a4paper, left=2.5cm, right=2.5cm, top=3.5cm, bottom=3cm]{geometry}
\usepackage{amssymb}
\usepackage{amsmath}
\usepackage{amscd}
\usepackage{latexsym}
\usepackage{graphicx}
\usepackage{ifthen}
\usepackage{epsfig}
\usepackage{setspace}
\usepackage{subcaption}
\usepackage{hyperref}
\usepackage{dsfont}
\usepackage{mathrsfs}
\usepackage{color}
\usepackage{cancel}
\usepackage{mathtools}
\usepackage{wrapfig}
\hypersetup{
 colorlinks=true,
 linkcolor=blue,
 citecolor=magenta,
}

\catcode`@=11
\renewcommand{\section}{\@startsection{section}{1}{0pt}{\medskipamount}
{\medskipamount}{\large\bf}}
\numberwithin{equation}{section}
\catcode`@=12

\newcommand{\C}{\mathds C}

\newcommand{\R}{\mathds R}
\newcommand{\g}{\frak g}
\newcommand{\h}{\frak h}
\newcommand{\m}{\frak m}
\newcommand{\Acal}{{\cal A}}
\newcommand{\Fcal}{{\cal F}}
\newcommand{\Ecal}{{\cal E}}
\newcommand{\Bcal}{{\cal B}}
\newcommand{\Tcal}{{\cal T}}
\newcommand{\Scal}{{\cal S}}

\newcommand{\Mcal}{{\cal M}}

\newcommand{\dd}{\text{d}}

\def\ep{\mathrm{e}}
\def\pa{\mbox{$\partial$}}
\def\diff{\mathrm{d}}
\def\tr{\mathrm{tr}}
\def\sfrac#1#2{{\textstyle\frac{#1}{#2}}}
\def\]{\right]}
\def\[{\left[}
\def\){\right)}
\def\({\left(}
\def\>{\rangle}
\def\<{\langle}
\def\+{\dagger}
\def\we{{\wedge}}
\def\={\ =\ }
\def\und{\quad\textrm{and}\quad}
\def\with{\quad\textrm{with}\quad}
\def\for{\quad\textrm{for}\quad}

\begin{document}

\title{\bf\huge 
Yang--Mills solutions on Minkowski space \\[4pt] 
via non-compact coset spaces}
\date{~}

\author{\phantom{.}\\[12pt]
\hspace{-.5cm}
{\Large Kaushlendra Kumar}, \ 
{\Large Olaf Lechtenfeld}, \ 
{\Large Gabriel Pican\c{c}o Costa}, \ 
{\Large Jona Röhrig}
\\[24pt]
{Institut f\"ur Theoretische Physik \&}\\ 
{Riemann Center for Geometry and Physics}\\
{Leibniz Universit\"at Hannover} \\ 
{Appelstra{\ss}e 2, 30167 Hannover, Germany}\\
[12pt] 
} 

\clearpage
\maketitle
\thispagestyle{empty}

\begin{abstract}
\noindent\large
We find a two-parameter family of solutions of the Yang--Mills equations 
for gauge group SO(1,3) on Minkowski space by foliating different parts of it 
with non-compact coset spaces with SO(1,3) isometry. 
The interior of the lightcone is foliated with hyperbolic space 
$H^3\cong \text{SO}(1,3)/\text{SO}(3)$, 
while the exterior of the lightcone employs de Sitter space 
dS$_3\cong \text{SO}(1,3)/\text{SO}(1,2)$. 
The lightcone itself is parametrized by SO(1,3)/ISO(2) in a nilpotent fashion.
Equivariant reduction of the SO(1,3) Yang--Mills system on the first two 
coset spaces yields a mechanical system with inverted double-well potential 
and the foliation parameter serving as an evolution parameter.
Its known analytic solutions are periodic or runaway except for the kink. 
On the lightcone, only the vacuum solution remains.
The constructed Yang--Mills field strength is singular across the lightcone
and of infinite action due to the noncompact cosets.
Its energy-momentum tensor takes a very simple form, with energy density
of opposite signs inside and outside the lightcone.
\end{abstract}

\newpage
\setcounter{page}{1} 

\section{Introduction and summary}
\noindent
Analytic solutions of the vacuum Yang--Mills equations (without Higgs fields) 
in Minkowski space are few and far between.
This holds in particular for compact gauge groups such as SU(2) 
and when finite energy or action is imposed
(see however~\cite{AFF,Luescher,Schechter} for examples).
The situation is less daunting when one allows for a noncompact gauge group,
in particular the Lorentz group.\footnote{
which of course features prominently in a gauge-theory formulation of General Relativity.}
Indeed, as we shall show, there exists a highly symmetric and geometrically distinguished
class of such field configurations.
The task of this paper is to present their geometric and algebraic construction
and to work out their basic properties.

The construction is based on the natural action of the Lorentz group on Minkowski space,
which foliates the latter into SO(1,3) orbits. One must distinguish four types of orbits:
The hyperbolic 3-space~$H^3$ in the future~${\cal T}_+$ or past~${\cal T}_-$ interior of 
the lightcone and the pseudo-Riemannian de Sitter space dS$_3$ in the exterior~${\cal S}$ 
of the lightcone are generic three-dimensional orbits coming in one-parameter families. 
The future and past lightcones~${\cal L}_\pm$ and the Minkowski origin are exceptional.
The generic orbits are reductive symmetric spaces $\text{SO}(1,3)/H$ 
with $H=\text{SO}(3)$ and $H=\text{SO}(1,2)$, respectively, 
and they are labelled by a foliation parameter~$u\in\R$.
Therefore, on the domains ${\cal T}_\pm$ and ${\cal S}$ we encounter 
SO(1,3) Yang--Mills theory on $\R\times\text{SO}(1,3)/H$.
Since we have taken the Yang--Mills structure group to agree with the isometry group
of our symmetric spaces, it is straightforward to write down the most general
SO(1,3)-symmetric gauge connection~$\Acal$ and find it dependent on a single real
function~$\phi(u)$. The Yang--Mills equations then translate to Newton's equation
for a particle in position~$\phi$ in an inverted double-well potential 
$V(\phi)=-\tfrac12(\phi^2{-}1)^2$. The two-dimensional family of solutions~$\phi(u)$ 
(parametrized by the double-well ``energy''~$\epsilon$ and a reference~$u_0$) 
then produce a family of classical Yang--Mills field configurations.
Given an explicit parametrization of the foliation, we can write down the
Yang--Mills connections and the field strength in Minkowski coordinates.

Here we perform this program and obtain explicit SO(1,3)-symmetric Yang--Mills fields 
inside and outside the Minkowski lightcone (of an arbitrary reference point) in terms of 
Minkowski coordinates~$x^\mu$ and the function $\phi\bigl(\tfrac12\ln|x{\cdot}x|\bigr)$,
where $x{\cdot}x=\eta_{\mu\nu}x^\mu x^\nu=\pm\ep^{2u}$ with Minkowski metric~$\eta$.
The action for either domain (${\cal T}_\pm$ or ${\cal S}$) is infinite due to the
infinite volume of the (noncompact) orbits.
The field strengths diverge as $|x{\cdot}x|^{-3/2}$ at the lightcone, 
and the energy-momentum tensor on ${\cal T}_\pm$ and on ${\cal S}$ takes the common form
\begin{equation} \label{EMtensor}
T_{\mu\nu} \= \frac{\epsilon}{g^2}\,
\frac{4\,x_\mu x_\nu - \eta_{\mu\nu}x{\cdot}x}{(x{\cdot}x)^3} \=
\pa^\rho S_{\rho\mu\nu} \qquad\textrm{with}\qquad
S_{\rho\mu\nu} \= \frac{\epsilon}{g^2}\,
\frac{ x_\rho \eta_{\mu\nu} - x_\mu \eta_{\rho\nu} }{(x{\cdot}x)^2}
\end{equation}
where $g$ denotes the Yang--Mills coupling constant.
Curiously, this energy-momentum tensor is of a pure ``improvement'' form,
which suggests the total energy and momentum to vanish on any spatial slice
provided a sufficient fall-off at spatial infinity. 
However, due to the singular behavior on the lightcone, 
the energy and momentum integrals each reduce to a divergent boundary term.
Matching of the field configurations on the two domains
for a solution covering the entire Minkowski space then requires
some regularization across the lightcone, where all densities change
sign due to the denominator in \eqref{EMtensor}.
A standard principal-value prescription does not suffice 
since the pole is of third order.\footnote{
A second-order pole will remain, although a fine-tuned principal-value 
recipe can remove all poles.}
Alternatively, additional degrees of freedom localized on the lightcone
may provide a source which compensates for the singularity in~\eqref{EMtensor}.

Finally, we investigate the non-reductive coset structure 
${\cal L}_\pm=\text{SO}(1,3)/\text{ISO}(2)$
of the lightcone itself. It is not a symmetric space but another subgroup generated by two
nilpotent (of degree 3) translations and one dilatation (which squares to a projector).
This provides a curious parametrization of the null hypersurface~${\cal L}_\pm$.

\newpage

\section{$G$-invariant Yang--Mills fields on $\R\times G/H$}
\noindent
Throughout this paper, we will discuss $G$-invariant Yang--Mills theories over some spacetime $M \subset \R^{1,3}$, where the gauge group $G$ also acts naturally on $M$. One can then use the orbit-stabilizer theorem to foliate $M$ by orbits of $G$, and then parametrize each orbit by the coordinates of some appropriate coset space. More specifically, the aforementioned theorem states that, $\forall x \in M$, there is a bijection between the orbit $G \cdot x$ and the quotient $G/G_x$, where $G_x=:H$ is the stabilizer of $x$. By foliating $M$ with orbits of $G$, one can then use coordinates of each coset space (together with the foliation parameter(s)) to also parameterize $M$.
Such coset coordinates are conveniently obtained from a parametrization of~$G$ by acting with~$G$ on some base vector~$x_0$, for each orbit.

For all the cases treated here we will be dealing with homogeneous spaces $G/H$ with 6-dimensional Lie groups $G$ and 3-dimensional stabilizer subgroups $H$. For reductive cosets, the Lie algebra $\g=\text{Lie}(G)$ splits into a 3-dimensional subalgebra $\h$ and its orthogonal complement $\m$ with respect to the Cartan--Killing metric.\footnote{In reductive homogeneous spaces, $\m$ remains invariant under the adjoint action of $H$, i.e., $h^{-1}\,\m\,h \subset \m,\ \forall\ h \in H$. This condition can also be written as $[\h,\m] \subset \m$.} This means that, for some specific basis of $\g$, the generators $\lbrace I_{\scriptscriptstyle{A}}\rbrace$ with structure constants $f_{{\scriptscriptstyle AB}}^{\quad {\scriptscriptstyle C}}$ satisfying 
\begin{equation}\label{LieAlg}
    [I_{\scriptscriptstyle{A}}, I_{\scriptscriptstyle{B}}] \= f_{{\scriptscriptstyle AB}}^{\quad {\scriptscriptstyle C}}\, I_{\scriptscriptstyle{C}}\ , \quad\with  A,B,C\ =\ 1,...,6\ ,
\end{equation}
inherit a likewise splitting:
\begin{equation} \label{split}
    \g \= \h \oplus \m\quad \implies\quad \lbrace I_{\scriptscriptstyle{A}}\rbrace \= \lbrace I_{i}\rbrace \cup \lbrace I_{a}\rbrace \quad\with a = 1,2,3\quad \text{and}\quad i = 4,5,6 .
\end{equation}
That is, $\{I_i\}$ is a basis of $\h$ and $\{I_a\}$ is a basis of $\m$. Furthermore, the commutation relations \eqref{LieAlg} for a reductive homogeneous space decompose as
\begin{equation} \label{LieAlgSplit}
    [I_i,I_j]\=f_{ij}^{\ \ k}I_k\ , \qquad [I_i,I_a]\=f_{ia}^{\ \ b}I_b\ , \,\und\, [I_a,I_b]\=f_{ab}^{\ \ i}I_i + f_{ab}^{\ \ c}I_c\ ,
\end{equation}
where the structure constants $f_{ab}^{\ \ c}=0$ for the special case of symmetric spaces,\footnote{Symmetric spaces are those reductive homogeneous spaces that additionally satisfy the condition $[\m,\m] \subset \h$.} which we will be mostly interested in. The Cartan--Killing metric is defined on $\g$ using its adjoint action on itself and can be written explicitly using the structure constants:
\begin{equation} \label{CKmetric}
    g_{{\scriptscriptstyle AB}} = - \tr_\text{ad}{\left(I_{\scriptscriptstyle{A}}\,I_{\scriptscriptstyle{B}}\right)} = f_{{\scriptscriptstyle AD}}^{\quad {\scriptscriptstyle C}}\,f_{{\scriptscriptstyle CB}}^{\quad {\scriptscriptstyle D}}\ ,
\end{equation}
where the trace is taken in the adjoint representation (this will be true for all trace operations in this paper). Note that $g_{{\scriptscriptstyle AB}}$ is positive-definite for compact Lie groups $G$ but is indefinite for the Lorentz group $\textrm{SO}(1,3)$ to be considered later on. 

To formulate the Yang--Mills gauge theory on $\R \times G/H$ we start with the principal bundle $P(G,G/H,\pi)$ with the structure group $G$ and canonical projection
\begin{equation}
    \pi: G \longrightarrow G/H\ ,\qquad g \mapsto g\cdot H\ .
\end{equation}
The Lie algebra $\g$ has an alternative formulation in terms of left-invariant vector fields $L_{\scriptscriptstyle A}$, satisfying commutation relations with the same structure constants as in \eqref{LieAlg}. One can obtain one-forms $\Tilde{e}^{\scriptscriptstyle A}$ dual to $L_{\scriptscriptstyle A}$ via the Maurer--Cartan prescription:
\begin{equation}
    g^{-1}\,\diff{g} \= \Tilde{e}^{\scriptscriptstyle A}\, I_{\scriptscriptstyle A}\ , \quad\for g \in G\ ,
\end{equation}
which can then be pulled back to the coset space $G/H$ using any local section $\sigma:\, G/H \supset U \longrightarrow G$ to obtain $e^{\scriptscriptstyle A} = \sigma^{*}\Tilde{e}^{\scriptscriptstyle A}$. For reductive cosets, these one-forms split into $\lbrace e^{\scriptscriptstyle A} \rbrace = \lbrace e^i \rbrace \cup \lbrace e^a \rbrace$ and satisfy (with proper normalization) the following structure equations consistent with \eqref{LieAlgSplit}:
\begin{equation} \label{StrucEqns}
    \diff e^a + f_{ib}^{\ \ a}\,e^i\wedge e^b \= 0 \qquad\und\qquad \diff e^i + \sfrac12 f_{jk}^{\ \ i}\,e^j\wedge e^k + \sfrac12 f_{ab}^{\ \ i}\,e^a\wedge e^b \= 0\ .
\end{equation}
Here, $e^i = e^i_a\,e^a$ are linearly dependent on the three $e^a$ on $G/H$, in terms of some real functions $e^i_a$. When the coset space is reductive, the set $\lbrace e^u{:=} \diff{u}, e^a \rbrace$, with the foliation parameter $u\in\R$, provides an orthonormal frame on the cotangent bundle $T^*(\R\times U)$. A generic gauge connection $\Acal$ and its curvature $\Fcal = \diff{\Acal} + \Acal\wedge\Acal$ can be expressed in this frame as
\begin{equation}
    \Acal \= \Acal_u\,e^u + \Acal_a\,e^a \qquad\implies\qquad \Fcal = \Fcal_{ua}\,e^u\wedge e^a + \sfrac12 \Fcal_{ab}\,e^a\wedge e^b\ ,
\end{equation}
where we set $\Acal_u{=}0$ using the ``temporal" gauge. Now expanding the gauge field in terms of the generators \eqref{split}, i.e. $\Acal_a = \Acal_a^i\,I_i + \Acal_a^b\,I_b$, and imposing $G$-invariance \cite{KZ92} yields~\footnote{The second relation can be written more concisely as $[I_i,\widetilde{\Acal}_a] = f_{ia}^{\ \ b}\widetilde{\Acal}_b$ for $\widetilde{\Acal}_a := \Acal_b^a\,I_a \in \m$.}
\begin{equation} \label{G-equiv}
    \Acal_a^i \= e^i_a \,\und\, \Acal_b^a=\Acal_b^a(u)\ , \with f_{ia}^{\ \ c}\Acal_b^a \= f_{ib}^{\ \ a}\Acal_a^c\ ,
\end{equation}
where for a symmetric space the remaining functions reduce to $\Acal^a_b(u) = \phi(u)\,\delta^a_b$ with a single real function~$\phi$. As a result, our $G$-invariant gauge field $\Acal$ is a deviation from the canonical $H$-connection $I_i\,e^i$:
\begin{equation} \label{ansatz}
    \Acal \= I_i\,e^i + \phi(u)\,I_a\,e^a\ .
\end{equation}
This yields, after some calculation involving \eqref{G-equiv} and \eqref{StrucEqns}, the components of the field strength $\Fcal$:
\begin{equation} \label{Fcal}
    \Fcal_{ua} \= \dot{\phi}\,I_a  \,\und\, \Fcal_{ab} \= (\phi^2{-}1)f_{ab}^{\ \ i}\,I_i\ , \quad\with \dot{\phi}:=\partial_u \phi\ ,
\end{equation}
which gives us the color-magnetic field $\Bcal_a = \sfrac12 \varepsilon_{abc}\Fcal_{bc} \in \h$ valued in the Lie subalgebra and the color-electric field $\Ecal_a = \Fcal_{au} \in \m$ valued in the subalgebra's orthogonal complement. The dynamics of $\phi(u)$ can then be extracted from the Yang--Mills equation by an extremization of the action:
\begin{equation}\label{YMaction}
    \delta\int \tr_\text{ad}{\left(\Fcal\wedge*\Fcal\right)} = 0 \qquad\implies\qquad \diff{*\Fcal} + \Acal\we{*\Fcal} - *\Fcal\we{\Acal} = 0\ .
\end{equation}

\section{A prototype: $\mathrm{SO}(4)/\mathrm{SO}(3) \cong S^3\ (\cong \mathrm{SU}(2)) $}
\noindent
In this section, we exemplify the previous discussion of a $G$-invariant Yang--Mills theory with the example of the compact group $G = \mathrm{SO}(4)$ acting on $M = \R^4$. For every $x \in \R^4 \backslash \{0\}$, the stabilizer subgroups $G_x$ are all identical, namely $\mathrm{SO}(3)=:H$. Therefore, every SO(4)-orbit (under left SO(3)-multiplication) is a (left) coset and, geometrically speaking, $\mathrm{SO}(4)/\mathrm{SO}(3)$ is the same as a round $3$-sphere~$S^3$. This allows us to foliate $\R^4\backslash \{0\}$ by $S^3$-slices labelled with radius $r\equiv\ep^u$ as the spatial foliation parameter. This becomes apparent with following maps ($\alpha,\beta=1,2,3,4$):
\begin{equation}\label{R4folitation}
\begin{aligned}
    \varphi &:~ \R\times S^3 \rightarrow \R^4\ , \quad (u,y^\alpha) \mapsto x^\alpha := \ep^u\,y^\alpha \qquad\textrm{with}\quad y{\cdot}y=1\ , \\
    \varphi^{-1} &:~ \R^4 \rightarrow \R\times S^3\ ,\quad x^\alpha \mapsto (u,y^{\alpha}) := \Bigl(\ln{\sqrt{x{\cdot}x}},\frac{x^\alpha}{\sqrt{x{\cdot}x}} \Bigr)\ ,
\end{aligned}
\end{equation}
where $x{\cdot}x:=\delta_{\alpha\beta}\,x^\alpha x^\beta$ and likewise for $y$. With this, the metric on $\R\times S^3$ -- with induced $S^3$-metric $\diff{\Omega}_{3}^2$ -- becomes conformal to not only $\R^4$, but also to its one-point compactification $S^4$ (of radius $\ell$; see \cite[Section 4]{ILP17(2)} for details) as well:
\begin{equation}\label{S4metric}
    \diff{s}^2 \= \ep^{2u}\bigl( \diff{u}^2 + \diff{\Omega}_{3}^2 \bigr) \= \frac{\ell^2}{(1{+}\cos\omega)^2}\Bigl( \diff{\omega}^2 + \sin^2\!\omega\,\bigl[\diff{\chi}^2 + \sin^2\!\chi\,(\diff{\theta}^2+\sin^2\!\theta\,\diff{\varphi}^2) \bigr] \Bigr)\ ,
\end{equation}
where $\ep^u=\ell\tan\frac{\omega}{2}$ so that $\omega,\chi,\theta\in[0,\pi]$ and $\varphi\in[0,2\pi]$ are canonical coordinates on $S^4$ and $(\chi,\theta,\varphi)$ parametrize the equatorial~$S^3$, whose line element $\diff{\Omega}_{3}^2$ sits in the square bracket. The canonical splitting \eqref{LieAlgSplit} of the Lie algebra $\g = \frak{so(4)}$ for this coset space is given by
\begin{equation}
    I_i \in \lbrace \Mcal_{23}, \Mcal_{31}, \Mcal_{12} \rbrace\quad \text{and} \quad I_a \in \lbrace \Mcal_{14}, \Mcal_{24}, \Mcal_{34} \rbrace\ ,\for \left( \Mcal_{\alpha\beta} \right)_{\gamma\delta} := \delta_{\alpha\delta}\,\delta_{\beta\gamma} - \delta_{\alpha\gamma}\,\delta_{\beta\delta}\ ,
\end{equation}
where $\alpha,\beta,\gamma,\delta = 1,2,3,4$. The corresponding structure constants are
\begin{equation}
    f_{ij}^{\ \ k} \= \varepsilon_{i-3\;j-3\;k-3}\ ,\quad f_{ia}^{\ \ b} \= \varepsilon_{i-3\;a\,b} \und   f_{ab}^{\ \ i} \= \varepsilon_{a\,b\,i-3}\ ,
\end{equation}
which produce the following Cartan--Killing metric \eqref{CKmetric}:
\begin{equation}
    g_{ij} \= f_{ik}^{\ \ l}\,f_{lj}^{\ \ k} + f_{ia}^{\ \ b}\,f_{bj}^{\ \ a} \= 4\,\delta_{ij}\ , \quad g_{ab} \= 2\,f_{ac}^{\ \ i}\,f_{ib}^{\ \ c} \= 4\,\delta_{ab}\ \und g_{ia} \= 0\ .
\end{equation}

We can identify the round sphere $S^3$ with coset space SO(4)/SO(3) as follows:
\begin{equation}
\begin{aligned}
    \alpha &:~ \mathrm{SO}(4)/\mathrm{SO}(3) \rightarrow S^3\ ,\quad [\Lambda] \mapsto y^\alpha := (\Lambda)^{\alpha}_{\;\;4}\ ,\\ 
    \alpha^{-1} &:~ S^3 \rightarrow \mathrm{SO}(4)/\mathrm{SO}(3)\ , \quad y^\alpha \mapsto [\Lambda]\ ,
\end{aligned}    
\end{equation}
where we define the equivalence class $[\Lambda]:=\{\widetilde{\Lambda}\in\mathrm{SO}(4):\widetilde{\Lambda}\sim\Lambda\}$ arising from the equivalence relation under right SO(3)-multiplication. The explicit form of a representative element $\Lambda$ of this class can be given by
\begin{equation}
\Lambda \= \begin{pmatrix} \mathds{1} + (\gamma{-}1)\frac{\pmb{\beta}\otimes\pmb{\beta}}{\pmb{\beta}^2} & \gamma\,\pmb{\beta} \\ -\gamma\,\pmb{\beta}^T & \gamma \end{pmatrix}\ , \quad\with
\beta^a = \frac{y^a}{y^4} \und \gamma = \frac{1}{\sqrt{1+\pmb{\beta}^2}} = y^4\ ,
\end{equation}
where $\pmb{\beta}^2=\delta_{ab}\,\beta^a\beta^b\ge0$. It is a straightfoward exercise to verify that the map $\alpha$ is well-defined. The representative element $\Lambda$ can simply be obtained by exponentiation with the coset generators $I_a\in \m$:
\begin{equation}\label{Texp}
    \Lambda \= \textrm{exp}(\eta^a\,I_a) \quad\with \beta^a = \sfrac{\eta^a}{\sqrt{\pmb{\eta}^2}}\tan{\sqrt{\pmb{\eta}^2}} \for \pmb{\eta}^2 = \delta_{ab}\,\eta^a\eta^b\ .
\end{equation}
We obtain Maurer--Cartan one-forms on $S^3$ with the representative element $\Lambda$ as follows
\begin{equation}
    \Lambda^{-1}\,\diff{\Lambda} \= e^a\,I_a + e^i\,I_i\ , \with e^a \= -\Bigl(\delta^{ab} + \frac{y^a\,y^b}{y^4(1{+}y^4)}\Bigr)\,\mathrm{d}y^b \und e^i \= \varepsilon_{a\,i-3\,b}\,\frac{y^a}{1{+}y^4}\,\mathrm{d}y^b\ ,
\end{equation}
and find that $e^a$ provide a local orthonormal frame on $S^3$, while $e^i$ are linearly dependent:
\begin{equation}\label{1formRel1}
    \diff{\Omega}_{3}^2 \= \delta_{ab}\,e^a\otimes e^b \,\und\, e^i \= e^i_a\,e^a\ , \with e^i_a \= \varepsilon_{a\,i-3\;b}\,\frac{y^b}{1{+}y^4}\ .
\end{equation}

\begin{wrapfigure}[8]{r}[1pt]{5cm}
\vspace{-30pt}
\centering
\includegraphics[width=4.5cm]{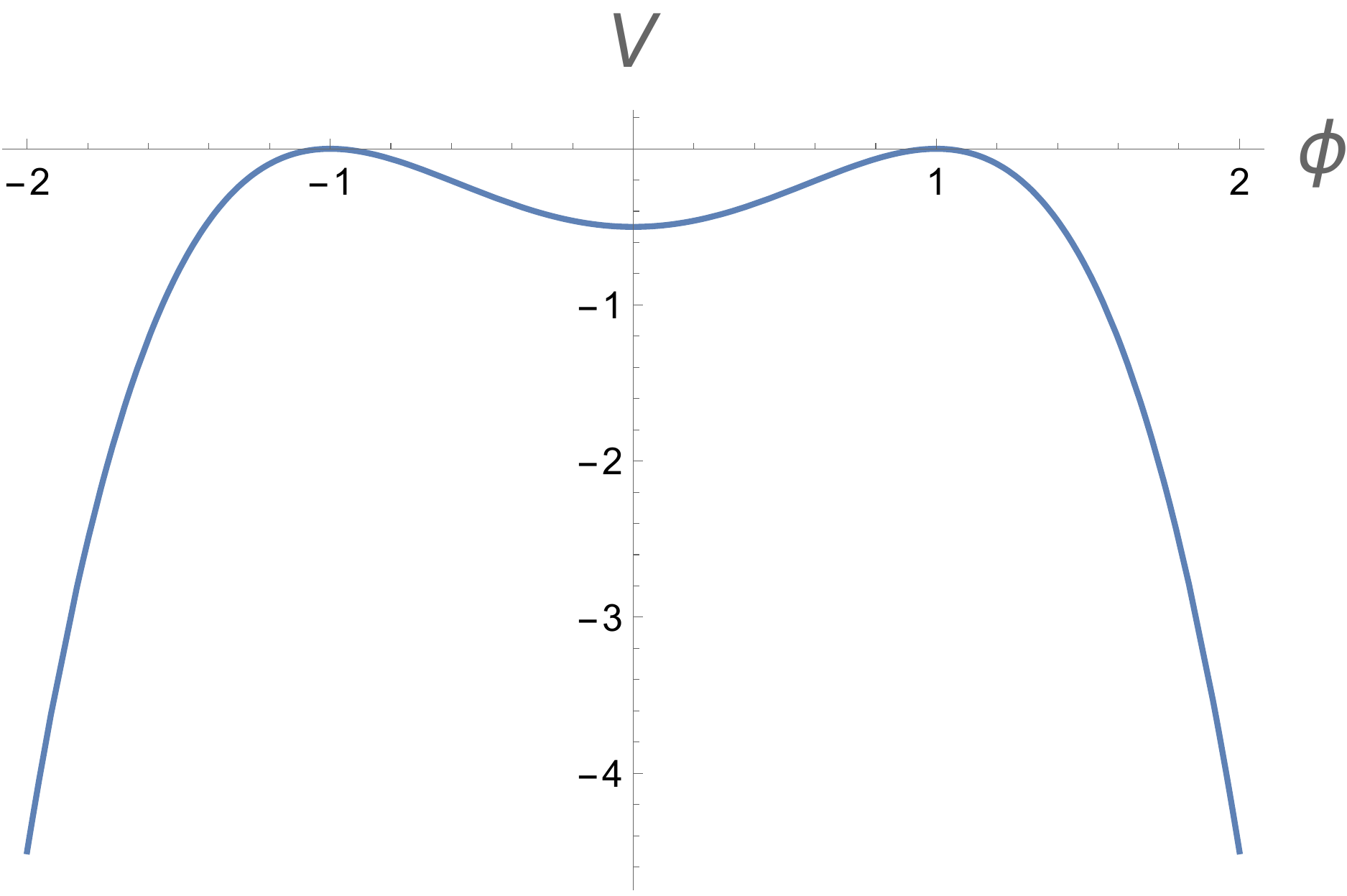}
\vspace{-5pt}
\caption{Plot of $V(\phi)$}
\label{invertedPotential}
\end{wrapfigure}

Using \eqref{ansatz} for $\R\times \mathrm{SO}(4)/\mathrm{SO}(3)$, the Yang--Mills equation is reduced to that of a mechanical particle under the influence of an inverted double-well potential $V(\phi)$ (see Figure \ref{invertedPotential})
\begin{equation}\label{Vphi}
    \ddot{\phi} \= 2\,\phi\,(\phi^2{-}1) \= - \frac{\partial V}{\partial \phi}\ ; \quad V(\phi) \= -\sfrac12 (\phi^2{-}1)^2\ ,
\end{equation}
whose solutions are known in terms of Jacobi elliptic functions. 

The equation of motion (\ref{Vphi}) also follows from the first-order differential equation arising from the self-duality condition in the Euclidean gauge theory, i.e.~instanton solutions, as also discussed in \cite{ILP17(2),IL08,ILPR09} and references therein. Similar features will also show up in Sections \ref{sec:interior} and \ref{sec:exterior} below. Solutions of the first-order differential equation are, consequently, special solutions of (\ref{Vphi}). The ``kink" solution \eqref{kinksoln} is one such example.

Now, using analytic continuation of $T$ in \eqref{S4metric} and conformal invariance of Yang--Mills theory in $4$ dimensions, we can pull these finite-energy finite-action solutions back to any conformally related spacetime. A prominent example here is the $4$-dimensional de Sitter space dS$_4 \ni (u,\chi,\theta,\varphi)$ of radius $\ell$ \cite{ILP17,ILP17(2),LU18}, which is also conformally related to the Minkowski space $\R^{1,3} \ni (t,x,y,z)$ \cite{LZ18}:
\begin{equation}
    \diff{s}_{\text{dS}_4}^2 \= \frac{\ell^2}{\cos^2{u}}\left( -\diff{u}^2 + \diff{\chi}^2 + \sin^2\!\chi\,(\diff{\theta}^2+\sin^2\!\theta\,\diff{\varphi}^2)\right)  \= \frac{\ell^2}{t^2} \left( -\diff{t}^2 + \diff{x}^2 + \diff{y}^2 + \diff{z}^2 \right)\ .
\end{equation}

\section{Interior of lightcone: $\mathrm{SO}(1,3)/\mathrm{SO}(3) \cong H^3$}
\label{sec:interior}
\noindent
From now onwards we work with the non-compact Lorentz group $G = \mathrm{SO}(1,3)$. In the following, we need to consider three different non-trivial stabilizer subgroups depending on the choice of a base vector $V \in \R^{1,3}$. The first subgroup is $\mathrm{SO}(3)$, which will be used to foliate (each half of) the interior of the lightcone. The second one is $\mathrm{SO}(1,2)$, which will be used to foliate the exterior of the lightcone. Lastly, the third one is the Euclidean group $\mathrm{E}(2)=\mathrm{ISO}(2)$, which can be used to parameterize (each half of) the lightcone itself. The chosen base vectors in the three cases are $(\pm \ep^u,0,0,0)$, $(0,0,0,\ep^u)$ and $(\pm \ep^u,0,0,\ep^u)$ respectively, for any $u\in \R$.

The interior $\Tcal$ of the lightcone may be foliated with two-sheeted hyperbolic space $H^{3} \cong \mathrm{SO}(4)/\mathrm{SO}(3)$ of constant curvature $-1$, as will be shown in what follows (see also \cite{LP2015}). It can be embedded in Minkowski space $\R^{1,3} \ni (y^\mu)$ as
\begin{equation} \label{H3}
    y{\cdot}y\equiv \eta_{\mu\nu}\,y^{\mu}\,y^{\nu} \= -1\ , \quad\with \eta_{\mu\nu} \= \textrm{diag}(-1,1,1,1)_{\mu\nu} \und \mu,\nu=0,1,2,3\ .
\end{equation}
The foliation of $\Tcal$ with $H^3$-sheets is then made explicit via the following maps:
\begin{equation}\label{Tfolitation}
\begin{aligned}
    \varphi_{_\Tcal} &:~ \mathds{R}\times H^3 \rightarrow \Tcal\ , \quad (u,y^\mu) \mapsto x^\mu := \ep^u\,y^\mu\ , \\
    \varphi_{_\Tcal}^{-1} &:~ \Tcal \rightarrow \mathds{R}\times H^3\ ,\quad x^\mu \mapsto (u,y^{\mu}) := \Bigl(\ln{\sqrt{|x{\cdot}x|}},\frac{x^\mu}{\sqrt{|x{\cdot}x|}} \Bigr)\ ,
\end{aligned}
\end{equation}
so that $\ep^u = \sqrt{|x{\cdot}x|}$. We will sometimes employ the conventional nomenclature: $x^0{=}t, x^1{=}x, x^2{=}y, x^3{=}z$ together with $\vec{x} := (x^1,x^2,x^3)$. The metric on the interior of the lightcone becomes conformal to that of a Lorentzian cylinder $\mathds{R}\times H^3$:
\begin{equation}\label{metricT}
    \diff{s}_{_\Tcal}^2 \= \ep^{2u}\left( -\diff{u}^2 + \diff{s}_{H^3}^2 \right)\ ,
\end{equation}
where $\diff{s}_{H^3}^2$ is the metric on $H^3$ induced from \eqref{H3}. The parameter $u$ here is temporal.

\begin{figure}[h!]
    \centering
    \includegraphics[width = 6cm, height = 6cm]{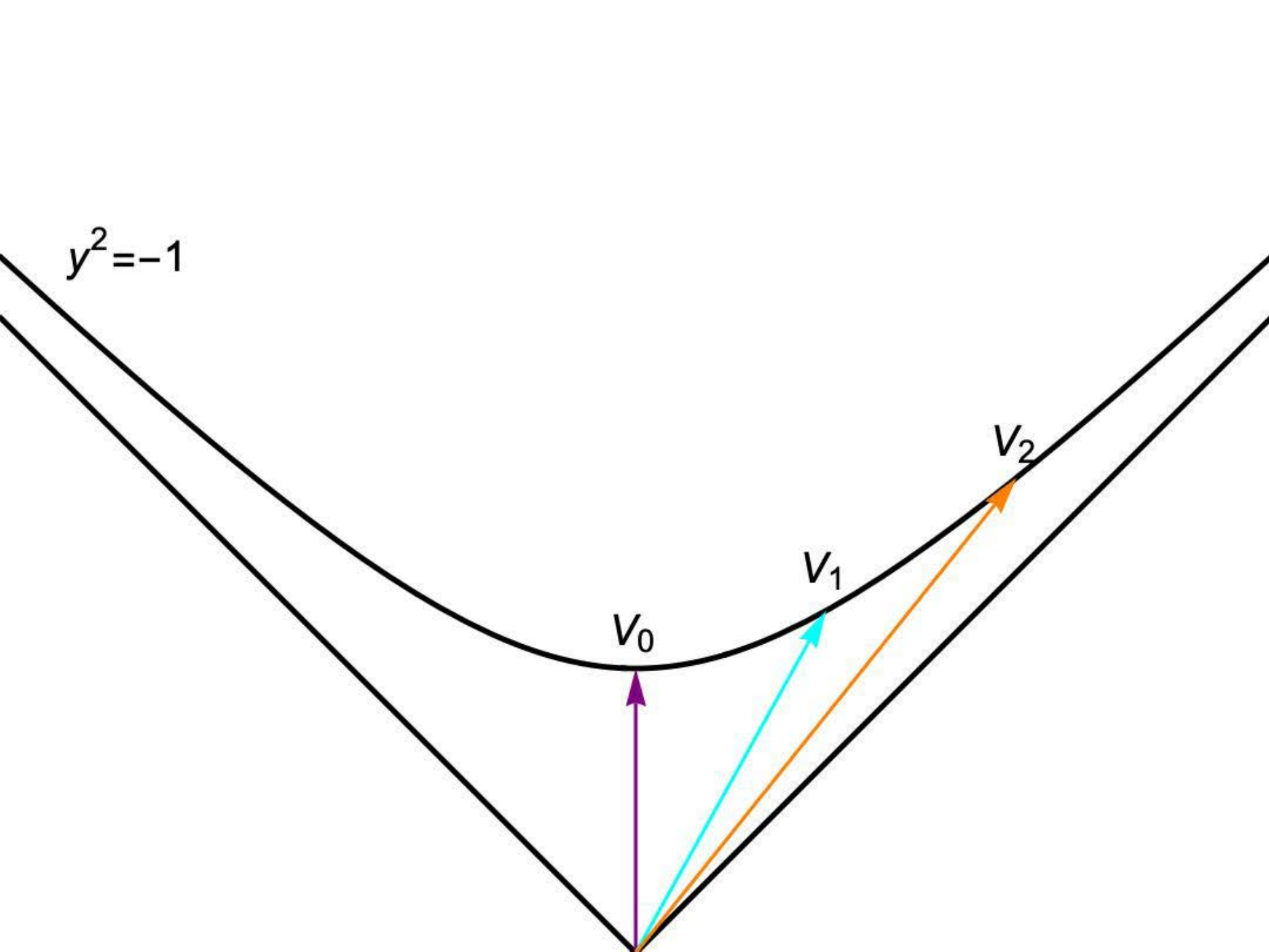}\hspace{2cm}
    \includegraphics[width = 6cm, height = 5cm]{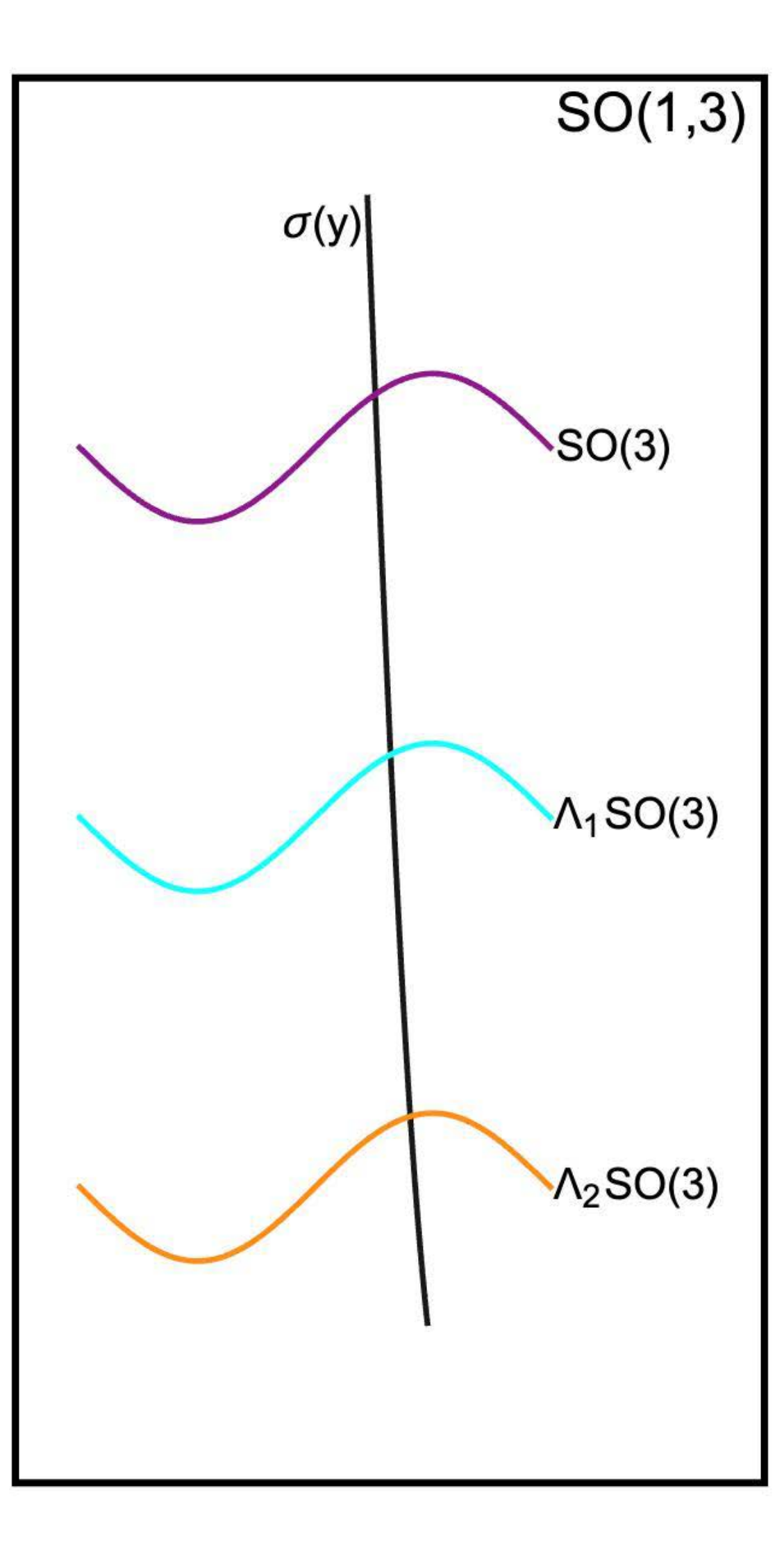}
    \caption{Left: Every vector $V_\alpha$ on $H^3$ can be brought to the temporal vector $V_0\sim (1,0,0,0)^\top$ with a unique boost $\Lambda_\alpha$, which determines its stabilizer as $\Lambda_\alpha\,\mathrm{SO}(3)\,\Lambda_\alpha^{\text -1}$. Right: Each vector belongs to some coset $\Lambda_\alpha\,\mathrm{SO}(3)$, and the choice of a representative~$\sigma$ in each coset yields $H^3$ as a 3-dimensional submanifold $\sigma(y)$ inside SO(1,3).} 
    \label{fig1}
\end{figure}

The canonical rotation ($J_a$) and boost ($K_a$) generators of the Lorentz group are given by
\begin{equation}
    J_1 {=} \begin{psmallmatrix} 0 & 0 & 0 & 0 \\ 0 & 0 & 0 & 0 \\ 0 & 0 & 0 & {\text -1} \\ 0 & 0 & 1 & 0  \end{psmallmatrix}, \ J_2 {=} \begin{psmallmatrix} 0 & 0 & 0 & 0 \\ 0 & 0 & 0 & 1 \\ 0 & 0 & 0 & 0 \\ 0 & {\text -1} & 0 & 0  \end{psmallmatrix}, \ J_3 {=} \begin{psmallmatrix} 0 & 0 & 0 & 0 \\ 0 & 0 & {\text -1} & 0 \\ 0 & 1 & 0 & 0 \\ 0 & 0 & 0 & 0  \end{psmallmatrix}, \
    K_1 {=} \begin{psmallmatrix} 0 & 1 & 0 & 0 \\ 1 & 0 & 0 & 0 \\ 0 & 0 & 0 & 0 \\ 0 & 0 & 0 & 0  \end{psmallmatrix}, \ K_2 {=} \begin{psmallmatrix} 0 & 0 & 1 & 0 \\ 0 & 0 & 0 & 0 \\ 1 & 0 & 0 & 0 \\ 0 & 0 & 0 & 0  \end{psmallmatrix}, \ K_3 {=} \begin{psmallmatrix} 0 & 0 & 0 & 1 \\ 0 & 0 & 0 & 0 \\ 0 & 0 & 0 & 0 \\ 1 & 0 & 0 & 0  \end{psmallmatrix},
\end{equation}
such that the splitting \eqref{split} for the coset space $\mathrm{SO}(1,3)/\mathrm{SO}(3)$ is realized as $I_i{=}J_{i-3}$ and $I_a {=} K_a$. The corresponding Lie algebra \eqref{LieAlgSplit} has the following structure coefficients:
\begin{equation}
    f_{ij}^{\ \ k} \= \varepsilon_{i-3\;j-3\;k-3}\ ,\qquad f_{ia}^{\ \ b} \= \varepsilon_{i-3\;a\,b}\ , \und   f_{ab}^{\ \ i} \= -\varepsilon_{a\,b\,i-3}\ .
\end{equation}
They produce an indefinite Cartan--Killing metric \eqref{CKmetric}, with
\begin{equation}
    g_{ij} \= f_{ik}^{\ \ l}\,f_{lj}^{\ \ k} + f_{ia}^{\ \ b}\,f_{bj}^{\ \ a} \= 4\,\delta_{i-3\;j-3}\ , \qquad g_{ab} \= 2\,f_{ac}^{\ \ i}\,f_{ib}^{\ \ c} \= -4\,\delta_{ab}\ ,\und g_{ia} \= 0\ .
\end{equation}
The hyperbolic space $H^3$ can be identified with the coset space $\mathrm{SO}(1,3)/\mathrm{SO}(3)$ (see Figure \ref{fig1} for an illustration). 
This is easily seen from the following maps,
\begin{equation}
\begin{aligned}
    \alpha_{_\Tcal} &:~ \mathrm{SO}(1,3)/\mathrm{SO}(3) \rightarrow H^3\ ,\quad [\Lambda_\Tcal] \mapsto y^\mu = (\Lambda_{\Tcal})^{\mu}_{\;\;0}\ ,\\ 
    \alpha_{_\Tcal}^{-1} &:~ H^3 \rightarrow \mathrm{SO}(1,3)/\mathrm{SO}(3)\ , \quad y^\mu \mapsto [\Lambda_\Tcal]\ ,
\end{aligned}    
\end{equation}
where $[\Lambda_\Tcal]=\{\Lambda\in\mathrm{SO}(1,3):\Lambda\sim\Lambda_\Tcal\}$ is the equivalence class (under right SO(3)-multiplication) of the representative
\begin{equation}\label{matrixT}
\Lambda_\Tcal \= \begin{pmatrix} \gamma & \gamma\,\pmb{\beta}^\top \\ \gamma\,\pmb{\beta} & \mathds{1} + (\gamma{-}1)\frac{\pmb{\beta}\otimes\pmb{\beta}}{\pmb{\beta}^2} \end{pmatrix}\ , \quad\with
\beta^a = \frac{y^a}{y^0}\ ,\quad \gamma = \frac{1}{\sqrt{1-\pmb{\beta}^2}} = y^0\ ,
\end{equation}
and $\pmb{\beta}^2=\delta_{ab}\,\beta^a\beta^b\ge0$. It can be checked that the map $\alpha_{_\Tcal}$ is well-defined. The representative element $\Lambda_\Tcal$ is nothing but a generic boost that can be obtained by exponentiation with the coset generators $I_a\in \m$:
\begin{equation}\label{Texp}
    \Lambda_\Tcal \= \textrm{exp}(\eta^a\,I_a)\ , \quad\with \beta^a = \sfrac{\eta^a}{\sqrt{\pmb{\eta}^2}}\tanh{\sqrt{\pmb{\eta}^2}}\ , \for \pmb{\eta}^2 = \delta_{ab}\,\eta^a\eta^b\ .
\end{equation}
The resulting Maurer--Cartan one-forms are
\begin{equation}
    \Lambda_\Tcal^{-1}\,\diff{\Lambda}_\Tcal \= e^a\,I_a + e^i\,I_i\ , \,\with\, e^a = \Bigl(\delta^{ab} - \frac{y^a\,y^b}{y^0(1{+}y^0)}\Bigr)\,\mathrm{d}y^b \und e^i = \varepsilon_{i-3\;a\,b}\,\frac{y^a}{1{+}y^0}\,\mathrm{d}y^b\ .
\end{equation}
Here, the $e^a$ provide a local orthonormal frame on $H^3$ while the $e^i$ are linearly dependent:
\begin{equation}\label{1formRel1}
    \diff{s}_{H^3}^2 \= \delta_{ab}\,e^a\otimes e^b \,\und\, e^i = e^i_a\,e^a\ , \with e^i_a = \varepsilon_{a\,i-3\;b}\,\frac{y^b}{1{+}y^0}\ .
\end{equation}
The Yang--Mills action simplifies to
\begin{equation}\label{YMaction}
S_{\text{YM}} \= -\frac{1}{4g^2}\int \tr_\text{ad}{\left(\Fcal\wedge*\Fcal\right)} \= \frac{6}{g^2}\int_{\mathds{R}\times H^3}\!\!\!\diff{\text{vol}}\ \bigl(\tfrac{1}{2}\dot{\phi}^2-V(\phi)\bigr)\ ,
\end{equation}
where $\diff{\text{vol}}=\sfrac{1}{3!}\varepsilon_{abc}\,\diff{u}\wedge e^a\wedge e^b\wedge e^c$ is the volume form and the potential $V(\phi) = -\sfrac12 (\phi^2{-}1)^2$ is the same as in \eqref{Vphi} and as depicted in Figure \ref{invertedPotential}. Therefore, this is the action of a mechanical particle in an {\em inverted\/} double-well potential $V(\phi)$, which yields the same equation of motion \eqref{Vphi} as before, i.e.
\begin{equation}\label{Teom}
    \ddot{\phi} \= - \frac{\partial V}{\partial \phi} \= 2\,\phi\,(\phi^2{-}1) \ .
\end{equation}
For $\epsilon\in[-\frac12,0]$ this equation admits bounded solutions in terms of, e.g., the Jacobi sine function sn:
\begin{equation} \label{YMsoln}
    \phi_{\epsilon,u_0}(u) \= f_{-}(\epsilon)\,\text{sn}\big(f_{+}(\epsilon)(u{-}u_0),k\big)\qquad\with\quad f_{\pm}(\epsilon) = \sqrt{1\pm\sqrt{-2\epsilon}}\ ,\quad k^2 = \frac{f_{-}(\epsilon)}{f_{+}(\epsilon)}\ ,
\end{equation}
which are characterized by the total mechanical energy $\epsilon = \sfrac12 \dot{\phi}^2 + V(\phi)$ and by a `time'-shift parameter $u_0$. Special cases, including the ``kink", of $\phi_{\epsilon,u_0}(u)$ are as follows,
\begin{equation}\label{kinksoln}
    \phi \= \begin{cases}
                 \ 0 \quad &\for\epsilon = -\sfrac12\ , \\
                 \ \tanh{(u{-}u_0)}\quad &\for\epsilon = 0\ ,\\
                 \ \pm 1 \quad &\for\epsilon = 0\ .\ 
            \end{cases}
\end{equation}
Moreover, we also have scattering solutions -- parameterized again with $\epsilon$ and $u_0$ -- given in terms of Jacobi functions cn and sn; for $\epsilon>0$ we have
\begin{equation}
    \phi_{\epsilon,u_0}(u) \= \frac{1}{\sqrt{2k^2 {-}1}}\frac{1+\textrm{cn}\left(\sfrac{2 }{\sqrt{2k^2 -1}}(u{-}u_0),k \right)}{\textrm{sn}\left(\sfrac{2 }{\sqrt{2k^2 -1}}(u{-}u_0),k \right)} \qquad\with \epsilon = -\frac{2k^2 (k^2 {-}1)}{(2k^2 {-}1)^2}\ ,
\end{equation}
while for $\epsilon<0$ we have
\begin{equation}
    \phi_{\epsilon,u_0}(u) \= \sqrt{\frac{2k^2 }{2k^2 {-}1}}\ \textrm{cn}\left(\sqrt{\sfrac{2}{1{-}2k^2 }}(u{-}u_0),k \right) \qquad\with \epsilon = -\frac{1}{2(2k^2 {-}1)^2}\ ,
\end{equation}
where $k^2>\sfrac12$ for both the cases.

It is a straightforward exercise to pull the orthonormal frame on $\R\times H^3$ back to $\Tcal$ with the map $\varphi_{_\Tcal}$ \eqref{Tfolitation} to obtain
\begin{equation}\label{Tframe}
    e^u := \diff{u} \= \frac{t\,\diff{t} - r\,\diff{r}} {t^2 - r^2} \qquad\und\qquad 
    e^a \= \frac{1}{|x|}\Bigl( \diff{x}^a - \frac{x^a}{|x|}\,\diff{t} + \frac{x^a}{|x|(|x| + t)}\,r\,\diff{r}\Bigr)\ ,
\end{equation}
where we have introduced the abbreviations
\begin{equation}
    |x| := \sqrt{|x{\cdot}x|}=\sqrt{|t^2{-}r^2|} \qquad\und\qquad 
    r := \sqrt{\vec{x}{\cdot}\vec{x}}\ .
\end{equation}
Like this, we can use $\varphi_{_\Tcal}$ to cast the SO(1,3)-invariant gauge field $\Acal \equiv A$ \eqref{ansatz} into a Minkowski one-form
\begin{equation}\label{AcalT}
    A \= \frac{1}{|x|}\left\{ \frac{\varepsilon_{ab}^{\ \ k-3}\,x^a}{|x|+t}\diff{x}^b\,I_k + \phi(x)\Bigl( \diff{x}^a - \frac{x^a}{|x|}\,\diff{t} + \frac{x^a}{|x|(|x| + t)}\,r\,\diff{r}\Bigr)I_a  \right\}\ ,
\end{equation}
where $\phi(x) {:=} \phi_{\epsilon,u_0}(u(x))$. One can then obtain the field strength $F=F_{\mu\nu}\,\mathrm{d}x^\mu\wedge\mathrm{d}x^\nu$ on $\Tcal$ by using the vierbein components $e^u = e^u_\mu\, \diff{x}^\mu$ and $e^a = e^a_\mu\,\diff{x}^\mu$ \eqref{Tframe} and the fields~\eqref{Fcal} on the cylinder. A straightforward computation then yields the color-electric $E_i:=F_{0i}$ and -magnetic $B_i:=\frac{1}{2}\varepsilon_{ijk}\,F_{jk}$ fields:
\begin{equation}\label{FcalT}
\begin{aligned}
    E_a &\= \frac{1}{|x|^3}\left\{\left(\phi^2{-}1\right)\varepsilon_{ab}^{\ \ i-3}\,x^b\,I_i -\Dot{\phi} \Bigl(t\,\delta^{ab} - \frac{x^a\,x^b}{|x|+t} \Bigr)I_b \right\}\ , \\
    B_a &\= -\frac{1}{|x|^3}\left\{\left(\phi^2{-}1\right)\Bigl(t\,\delta^{a\,i-3} - \frac{x^a\,x^{i-3}}{|x|+t}\Bigr)I_i +\Dot{\phi}\, \varepsilon_{ab}^{\ \ c}\,x^b\,I_c \right\}\ .
\end{aligned}
\end{equation}
Interestingly, the structure of these fields demonstrates the presence of color-electromagnetic duality: $E_a \rightarrow B_a$ and $B_a \rightarrow -E_a$ leaves the expressions invariant, provided we interchange $\Dot\phi\leftrightarrow(\phi^2{-}1)$ as well as the coset and Lie-subalgebra generators $I_i \rightarrow I_a$ and $I_a \rightarrow -I_i$, followed by index adjustment. Furthermore, both the gauge field $\Acal$ \eqref{AcalT} as well as the electric $E_i$ and magnetic $B_i$ fields \eqref{FcalT} are singular at the lightcone~$t{=}{\pm}r$. The expression for the stress-energy tensor
\begin{equation} \label{SEtensor}
    T_{\mu\nu} \= -\tfrac{1}{2g^2}\, \tr_\text{ad}{\left( F_{\mu\alpha}\,F_{\nu\beta}\,\eta^{\alpha\beta} - \sfrac14 \eta_{\mu\nu} F^2 \right)}\ , \,\with F^2 = F_{\mu\nu}\,F^{\mu\nu}\ ,
\end{equation}
of these Yang--Mills fields is straightforwardly computed to yield
\begin{equation}\label{stressEnergy}
    T \= \frac{\epsilon}{g^2(r^2{-}t^2)^3} \begin{pmatrix} 3t^2{+}r^2 & -4tx & -4ty & -4tz \\ -4tx & t^2{+}4x^2{-}r^2 & 4xy & 4xz \\ -4ty  & 4xy & t^2{+}4y^2{-}r^2 & 4yz \\ -4tz & 4xz & 4yz & t^2{+}4z^2{-}r^2 \end{pmatrix}\ .
\end{equation}
As expected, it has a vanishing trace, but the lightcone singularity present in the fields \eqref{FcalT} shows up here as well.

\section{Exterior of lightcone: $\mathrm{SO}(1,3)/\mathrm{SO}(1,2) \cong \diff{\mathrm{S}}_3$}
\label{sec:exterior}
\noindent
Let us now study the foliation of the exterior of the lightcone, denoted here as $\Scal$. By choosing the base vector $V_0\sim(0,0,0,1)^\top$, we single out the 3-direction and reveal the stabilizer as $\mathrm{SO}(1,2)$. The exterior of the lightcone can then be foliated by copies of $\mathrm{SO}(1,3)/\mathrm{SO}(1,2) \cong \text{dS}_3$, i.e.~3-dimensional de Sitter space. This can be embedded in Minkowski space $\R^{1,3}$ as
\begin{equation}\label{dS3}
    y\cdot y \equiv \eta_{\mu\nu}\,y^\mu\,y^\nu \= 1\ . 
\end{equation}
The foliation of $\Scal$ is then achieved, analogous to the $\Tcal$ case, using the maps
\begin{equation}\label{Sfolitation}
\begin{aligned}
    \varphi_{_\Scal} &:~ \R\times\text{dS}_3 \rightarrow \Scal\ , \quad (u,y^\mu) \mapsto x^\mu := \ep^u\,y^\mu \ , \\
    \varphi_{_\Scal}^{-1} &:~ \Scal \rightarrow \R\times\text{dS}_3\ ,\quad x^\mu \mapsto (u,y^\mu) := \Bigl(\ln{\sqrt{|x{\cdot}x|}},\frac{x^\mu}{\sqrt{|x{\cdot}x|}} \Bigr)\ ,
    \end{aligned}
\end{equation}
so that $\ep^u = \sqrt{|x{\cdot}x|}$. The metric on $\Scal$ becomes conformal to that on the cylinder $\R\times\text{dS}_3$ using $\varphi_{_\Scal}$:
\begin{equation}\label{metricS}
    \diff{s}_{_\Scal}^2 \= \ep^{2u}\left( \diff{u}^2 + \diff{s}_{\text{dS}_3}^2 \right)\ ,
\end{equation}
where $\diff{s}_{\text{dS}_3}^2$ is the metric on $\text{dS}_3$ induced from \eqref{dS3}, and the parameter $u$ is spatial. 

For the coset $\mathrm{SO}(1,3)/\mathrm{SO}(1,2)$ associated with the base vector $(0,0,0,1)^\top$, the splitting \eqref{split} is realized by
\begin{equation}
    I_i \in \lbrace K_1, K_2, J_3 \rbrace
    \,\und\,
    I_a \in \lbrace J_1, J_2, K_3 \rbrace \ .
\end{equation}
The structure coefficients for this Lie algebra \eqref{LieAlgSplit} are
\begin{equation}
    f_{ij}^{\ \ k} \= \varepsilon_{i-3\;j-3\;k-3}\,(1{-}2\,\delta_{k6})\ ,\qquad f_{ia}^{\ \ b} \= \varepsilon_{i-3\;a\,b}\,(1{-}2\,\delta_{a3}) \,\und\,  f_{ab}^{\ \ i} \= \varepsilon_{a\,b\,i-3}\ ,
\end{equation}
where the indices for the terms inside the bracket are not summed over. These structure coefficients produce the following
Cartan--Killing metric \eqref{CKmetric},
\begin{equation}
    g_{ij} \= 4\,\begin{psmallmatrix} -1 & 0 & 0 \\ 0 & -1 & 0 \\ 0 & 0 & 1  \end{psmallmatrix}_{i-3\;j-3}\ , \qquad g_{ab} \= 4\,\begin{psmallmatrix} 1 & 0 & 0 \\ 0 & 1 & 0 \\ 0 & 0 & {\text -1}  \end{psmallmatrix}_{ab}\ , \;\und\; g_{ia} \= 0\ .
\end{equation}
The following maps illustrate the equivalence between $\text{dS}_3$ and $\mathrm{SO}(1,3)/\mathrm{SO}(1,2)$:
\begin{equation}
\begin{aligned}
    \alpha_{_\Scal} &:~ \mathrm{SO}(1,3)/\mathrm{SO}(1,2) \rightarrow \diff{\text{S}}_3\ ,\quad [\Lambda_\Scal] \mapsto y^\mu := (\Lambda_\Scal)^{\mu}_{\;\;3}\ , \\
    \alpha_{_\Scal}^{-1} &:~ \diff{\text{S}}_3 \rightarrow \mathrm{SO}(1,3)/\mathrm{SO}(1,2)\ , \quad y^\mu \mapsto [\Lambda_\Scal]\ ,
\end{aligned}    
\end{equation}
where the representative element $\Lambda_\Scal$ of each SO(1,2) coset~$[\Lambda_\Scal]$ is defined as follows:
\begin{equation}\label{matrixS}
    \Lambda_\Scal = \begin{pmatrix} 1 {+} (\gamma{-}1)\frac{\beta_1^2}{\pmb{\beta}^2} & -(\gamma{-}1)\frac{\beta_1\beta_2}{\pmb{\beta}^2} & -(\gamma{-}1)\frac{\beta_1\beta_3}{\pmb{\beta}^2} & \beta_1\gamma \\ (\gamma{-}1)\frac{\beta_1\beta_2}{\pmb{\beta}^2} & 1 {-} (\gamma{-}1)\frac{\beta_2^2}{\pmb{\beta}^2} & -(\gamma{-}1)\frac{\beta_2\beta_3}{\pmb{\beta}^2} & \beta_2\gamma \\ (\gamma{-}1)\frac{\beta_1\beta_3}{\pmb{\beta}^2} & -(\gamma{-}1)\frac{\beta_2\beta_3}{\pmb{\beta}^2} & 1 {-} (\gamma{-}1)\frac{\beta_3^2}{\pmb{\beta}^2} & \beta_3\gamma \\ \beta_1\gamma & -\beta_2\gamma & -\beta_3\gamma & \gamma \end{pmatrix}, \with y^3\begin{pmatrix}\beta_1 \\ \beta_2 \\ \beta_3 \end{pmatrix} {=} \begin{pmatrix} y^0 \\ y^1 \\ y^2 \end{pmatrix} ,\ \gamma {=} \sfrac{1}{\sqrt{1{-}\pmb{\beta}^2}} {=} y^4\ ,
\end{equation}
but now $\pmb{\beta}^2 = -\eta^{ab}\,\beta_a\beta_b\ge0$, where the $3$-dimensional Minkowski metric $\eta_{ab} = \textrm{diag}(-1, 1, 1)_{ab}$ has its origin in the fact that the stabilizer $\mathrm{SO}(1,2)$ is nothing but the isometry group of~$\R^{1,2}$. As in the previous section, it can be checked that the map $\alpha_{_\Scal}$ is well-defined.
One can also obtain $\Lambda_\Scal$, in analogy to the previous case \eqref{Texp}, by exponentiating the coset generators $I_a\in \m$ with parameters $\kappa_a$:
\begin{equation}
    \Lambda_\Scal \= \textrm{exp}(-\kappa_3\,J_1 {+} \kappa_2\,J_2 {+} \kappa_1\,K_3) \quad\with \beta_a = \frac{\kappa_a}{\sqrt{\pmb{\kappa}^2}}\tanh{\sqrt{\pmb{\kappa}^2}} \for \pmb{\kappa}^2 = -\eta^{ab}\kappa_a\kappa_b \ .
\end{equation}
The parameter $\pmb{\kappa}^2$ can also be negative here and the expression above still holds, turning $\tanh$ into $\tan$. 
We obtain the following Maurer--Cartan one-forms:
\begin{equation}\label{1formS}
    \Lambda_\Scal^{-1}\,\diff{\Lambda}_\Scal \= e^a\,I_a + e^i\,I_i \with e^a = \diff{y}^{3-a} - \frac{y^{3-a}}{1{+}y^3}\,\diff{y}^3 \und e^i = -\varepsilon_{i-3\;a\,b}\,\frac{y^{3-a}}{1{+}y^3}\,\diff{y}^{3-b}\ .
\end{equation}
The one-forms $e^a$ provide a local orthonormal frame on $\diff{\text{S}}_3$ while the $e^i$ are linearly dependent:
\begin{equation}\label{1formRel2}
    \diff{s}_{\diff{\text{S}}_3}^2 \= \eta_{ab}\,e^a\otimes e^b \,\und\, e^i = e^i_a\,e^a\ , \with e^i_a = \varepsilon_{i-3\;a\,b}\,\frac{y^{3-b}}{1{+}y^3}\ ,
\end{equation}
such that the metric on the cylinder $\R\times\text{dS}_3$ \eqref{metricS} becomes
\begin{equation}
    \diff{s}_{_\Scal}^2 \= e^u\otimes e^u + \eta_{ab}\, e^a\otimes e^b\ .
\end{equation}
The Yang--Mills action on $\R\times\text{dS}_3$ comes out to be
\begin{equation}
    S_{\text{YM}} \= \frac{2}{g^2}\int_{\mathds{R}\times\diff{\text{S}}_3}\!\!\!\!\diff{\text{vol}}\ \bigl(\tfrac{1}{2}\dot{\phi}^2-V(\phi)\bigr)\ , \with V(\phi) \= -\sfrac12 (\phi^2{-}1)^2,
\end{equation}
where the volume form $\diff{\text{vol}}=\sfrac{1}{3!}\varepsilon_{abc}\,\diff{u}\wedge e^a\wedge e^b\wedge e^c$ depends on~\eqref{1formS}. 
Hence, we encounter the same equation of motion~\eqref{Teom} as before and, therefore,
the generic solution \eqref{YMsoln} coupled with \eqref{Fcal} applies here as well, albeit with a catch that electric $\Ecal_a$ and magnetic $\Bcal_a$ fields are valued in other spaces. 

Pulling the local orthonormal frame on $\R\times\text{dS}_3$ back to $\Scal$ with the map $\varphi_{_\Scal}$ \eqref{Sfolitation}, we obtain
\begin{equation}
    e^u := \diff{u} \= \frac{ r\,\diff{r} - t\,\diff{t}}{r^2 - t^2}\,\und\,
    e^a \= \frac{1}{|x|}\Bigl(\diff{x}^{3-a} - \frac{x^{3-a}}{|x|}\,\diff{z} - \frac{\eta_{bc}\,x^{3-a}\,x^{b-1}}{|x|(|x|+z)}\,\diff{x}^{c-1} \Bigr)\ ,
\end{equation}
which gives us the vierbein components $e^u {=} e^u_\mu\, \diff{x}^\mu$ and $e^a {=} e^a_\mu\,\diff{x}^\mu$. As before, we obtain the gauge potential on the cylinder $\R\times\textrm{dS}_3$,
\begin{equation}
    A \= \frac{1}{|x|}\bigg\{ \frac{\varepsilon^{i{-}3}_{\quad\ c\,b}\,x^{3-b}}{|x|+z}\,\diff{x}^{3-c}\,I_i + \phi(x)\Bigl(\diff{x}^{3-a} - \frac{x^{3-a}}{|x|}\,\diff{z} - \frac{\eta_{bc}\,x^{3-a}\,x^{b-1}}{|x|(|x|+z)} \,\diff{x}^{c-1} \Bigr)I_a\bigg\}\ ,
\end{equation}
which yields the following color-electric $E_i$ and -magnetic $B_i$ fields:
\begin{equation}
   \begin{aligned}
    E_1 &\= \frac{1}{|x|^3}\left[\dot{\phi}\, (I_2\,t+I_3\,x) + \left(\phi^2{-}1\right) \Bigl\{ -\frac{y}{|x| + z} (I_6\,t - I_5\,x - I_4\,y)
   +I_4\,z \Bigr\}\right]\ ,\\
   E_2 &\= \frac{1}{|x|^3}\left[\dot{\phi}\, (I_1\,t+I_3\,y) + \left(\phi^2{-}1\right) \Bigl\{ \frac{x}{|x| + z} (I_6\,t - I_5\,x - I_4\,y )
   - I_5\,z \Bigr\}\right]\ ,\\
   E_3 &\= \frac{1}{|x|^3}\left[\dot{\phi}\,\Bigl\{ -\frac{t}{|x| + z} (I_3\,t + I_1\,y + I_2\,x)
   +I_3\,z \Bigr\} - \left(\phi^2{-}1\right) (I_4\,x - I_5\,y) \right]\ ,\\
   B_1 &\= \frac{1}{|x|^3}\left[-\dot{\phi}\,\Bigl\{ \frac{y}{|x| + z} (I_3\,t + I_1\,y + I_2\,x)
   +I_1\,z \Bigr\} + \left(\phi^2{-}1\right) (I_5\,t - I_6\,x) \right]\ ,\\
   B_2 &\= \frac{1}{|x|^3}\left[\dot{\phi}\,\Bigl\{ \frac{x}{|x| + z} (I_3\,t + I_1\,y + I_2\,x)
   +I_2\,z \Bigr\} - \left(\phi^2{-}1\right) (I_6\,y - I_4\,t) \right]\ ,\\
   B_3 &\= \frac{1}{|x|^3}\left[\dot{\phi}\, (I_1\,x - I_2\,y) + \left(\phi^2{-}1\right) \Bigl\{ \frac{t}{|x| + z} (I_6\,t - I_5\,x - I_4\,y )
   - I_6\,z \Bigr\}\right]\ .
   \end{aligned}
\end{equation}

The resulting stress-energy tensor \eqref{SEtensor} comes out to be the same as on $\Tcal$, i.e.~\eqref{stressEnergy}, or
\begin{equation} \label{TandS}
    T_{\mu\nu} \= \pa^\rho S_{\rho\mu\nu} \qquad\with\qquad
    S_{\rho\mu\nu} \= \frac{\epsilon}{g^2}\,
\frac{ x_\rho \eta_{\mu\nu} - x_\mu \eta_{\rho\nu} }{(x{\cdot}x)^2}\ ,
\end{equation}
where the improvement term $(\tilde{S}_\rho)_{\mu\nu} := \sfrac{g^2(x{\cdot}x)^2}{\epsilon} S_{\rho\mu\nu}$ takes the following explicit form
\begin{equation}
    \tilde{S}_0 \= \begin{psmallmatrix} 
    0 & 0 & 0 & 0 \\
    x & -t & 0 & 0 \\
    y & 0 & -t & 0 \\
    z & 0 & 0 & -t
    \end{psmallmatrix}\ ,\quad \tilde{S}_1 \= \begin{psmallmatrix} 
    -x & t & 0 & 0 \\
    0 & 0 & 0 & 0 \\
    0 & -y & x & 0 \\
    0 & -z & 0 & x
    \end{psmallmatrix}\ ,\quad \tilde{S}_2 \= \begin{psmallmatrix} 
    -y & 0 & t & 0 \\
    0 & y & -x & 0 \\
    0 & 0 & 0 & 0 \\
    0 & 0 & -z & y
    \end{psmallmatrix}\ ,\quad \tilde{S}_3 \= \begin{psmallmatrix} 
    -z & 0 & 0 & t \\
    0 & z & 0 & -x \\
    0 & 0 & z & -y \\
    0 & 0 & 0 & 0
    \end{psmallmatrix}\ .
\end{equation}
It is tempting to combine the two stress tensors inside and outside the lightcone 
to a single expression valid on all Minkowski spacetime. The price is the singularity
on the lightcone, however. An attempt to regularize the latter is
\begin{equation} \label{EMreg}
    S^{\textrm{reg}}_{\rho\mu\nu} \= \frac{\epsilon}{g^2}\,
\frac{ x_\rho \eta_{\mu\nu} - x_\mu \eta_{\rho\nu} }{(x{\cdot}x+\delta)^2}
\qquad\Rightarrow\qquad
T^{\textrm{reg}}_{\mu\nu} \= \frac{\epsilon}{g^2}\,
\frac{4\,x_\mu x_\nu - \eta_{\mu\nu}x{\cdot}x + 3\,\delta\,\eta_{\mu\nu}}{(x{\cdot}x+\delta)^3}
\end{equation}
which, as a nonsingular improvement term, will give rise to vanishing energy and momenta 
for any finite value of the regularization parameter~$\delta$ 
(the fall-off at spatial infinity is fast enough).
Alternatively, shifting directly only the denominator of $T_{\mu\nu}$ in \eqref{TandS} via 
$x{\cdot}x\mapsto x{\cdot}x+\delta$ will yield a proper regular energy-momentum tensor,
so that, by equivalence under adding the improvement \eqref{EMreg}, we obtain that
\begin{equation}
    T^\delta_{\mu\nu} \= \frac{\epsilon}{g^2}\,
\frac{4\,x_\mu x_\nu - \eta_{\mu\nu}x{\cdot}x}{(x{\cdot}x+\delta)^3}
\ \sim\ 
\frac{\epsilon}{g^2}\,
\frac{-3\,\delta\,\eta_{\mu\nu}}{(x{\cdot}x+\delta)^3}
\end{equation}
provides a candidate for a regular energy-momentum tensor in the entire Minkowski spacetime.
Note that the latter expression vanishes for $\delta\to0$.

\section{Null hypersurface: $\mathrm{SO}(1,3)/\mathrm{ISO}(2) \cong {\cal L}_+$}
\label{sec:lightcone}
\noindent
The stabilizer subgroup $H \subset G$ associated with a base vector in the lightcone is not so straightforward to see as in the previous cases, but it can be easily computed using the double cover of the Lorentz group, $\mathrm{SL}(2,\C)$, and its action on the vector space of $2{\times}2$ Hermitian matrices, which is isomorphic to $\R^{1,3}$ (see for example \cite{Woit18}). 

The stabilizer subgroup $H$ in this case is the Euclidean group $\mathrm{E}(2) = \mathrm{ISO}(2)$ generated by two translations and one rotation. The subalgebra $\frak{h}$ is spanned by 
$\{I_4,I_5,I_6\}=\{P_3,P_4,J_3\}$, with
\begin{equation}
    P_3 {:=} K_1{-}J_2 \= \begin{psmallmatrix} 0 & 1 & 0 & 0 \\ 1 & 0 & 0 & {\text -1} \\ 0 & 0 & 0 & 0 \\ 0 & 1 & 0 & 0  \end{psmallmatrix}\ ,\qquad P_4 {:=} K_2 {+} J_1 \= \begin{psmallmatrix} 0 & 0 & 1 & 0 \\ 0 & 0 & 0 & 0 \\ 1 & 0 & 0 & {\text -1} \\ 0 & 0 & 1 & 0  \end{psmallmatrix}\ , \,\und\, J_3 \= \begin{psmallmatrix} 0 & 0 & 0 & 0 \\ 0 & 0 & {\text -1} & 0 \\ 0 & 1 & 0 & 0 \\ 0 & 0 & 0 & 0  \end{psmallmatrix}\ .
\end{equation}
Here again, the algebra $\frak{g}$ can be decomposed into $\frak{h}\oplus\frak{m}$, with $\frak{m}$ generating the coset $G/H$, but this coset is not reductive as in the previous cases. Indeed, $P_3$ and $P_4$ are orthogonal to themselves with respect to the Cartan--Killing metric \eqref{CKmetric}, so there is no orthogonal complement to $\frak{h}$.
Moreover, $\frak{m}$ also forms a subalgebra of $\frak{g}$ (see also \cite{BGH22}), spanned by
$\{I_1,I_2,I_3\}=\{P_1,P_2,K_3\}$, with
\begin{equation}
    P_1 {:=} K_1{+}J_2 \= \begin{psmallmatrix} 0 & 1 & 0 & 0 \\ 1 & 0 & 0 & 1 \\ 0 & 0 & 0 & 0 \\ 0 & {\text -1} & 0 & 0  \end{psmallmatrix}\ ,\qquad P_2 {:=} K_2 {-} J_1 \= \begin{psmallmatrix} 0 & 0 & 1 & 0 \\ 0 & 0 & 0 & 0 \\ 1 & 0 & 0 & 1 \\ 0 & 0 & {\text -1} & 0  \end{psmallmatrix}\ , \,\und\, K_3 \= \begin{psmallmatrix} 0 & 0 & 0 & 1 \\ 0 & 0 & 0 & 0 \\ 0 & 0 & 0 & 0 \\ 1 & 0 & 0 & 0  \end{psmallmatrix}\ .
\end{equation}
$P_1$ and $P_2$ act again as generators of translations, while $K_3$ is a generator of dilations. 
The subalgebras $\frak{h}$, spanned by $\{ I_i \}$, and $\frak{m}$, spanned by $\{ I_a \}$, are given by
\begin{equation}
        [I_4,I_6] = -I_5\ ,\quad [I_5,I_6] = I_4\ ,\quad [I_4,I_5] = 0 \und [I_1,I_3] = I_1\ ,\quad [I_2,I_3] = I_2\ ,\quad [I_1,I_2] = 0  \ ,
\end{equation}
respectively, while their non-orthogonality is demonstrated by the mixed commutators
\begin{equation}
\begin{aligned}
    [I_1,I_4] &\= -2\,I_3\ ,\quad [I_1,I_5] \= -2\,I_6\ ,\quad [I_2,I_4] \= 2\,I_6\ ,\quad [I_2,I_5] \= -2\,I_3\ ,\\
    [I_3,I_4] &\= I_4\ ,\quad [I_3,I_5] \= I_5\ ,\quad [I_1,I_6] \= I_2\ ,\quad [I_2,I_6] \= I_1\ ,\quad  [I_3,I_6] \= 0\ .
\end{aligned}
\end{equation}
The algebra spanned by $\frak{m}$ is known as a type V algebra in Bianchi's classification of 3-dimensional real Lie algebras, or as a $\frak{g}_{3.3}$ algebra in Mubarakzyanov's classification.

If we had chosen any base vector proportional to $(-1,0,0,1)^\top$, then the splitting would be realized by
\begin{equation}
    I_i \in \lbrace P_1, P_2, J_3 \rbrace \,\und\, I_a \in \lbrace P_3, P_4, K_3 \rbrace\ ,
\end{equation}
and we would use the group generated by the new $\frak{m}$ to parametrize the past half of the lightcone ${\cal L}_-$ in the same way as the one described below for parametrizing ${\cal L}_+$.

Let us map the coset space into the future half of the lightcone using the base vector $(1,0,0,1)^\top$ and the map
\begin{equation}\label{maplightcone}
\begin{aligned}
    \alpha_{{\cal L}_+} &:~ \mathrm{SO}(1,3)/\mathrm{ISO}(2) \supset \exp(\frak{m}) \rightarrow {\cal L}_+, \quad [\Lambda_{{\cal L}_+}] \mapsto y^\mu = (\Lambda_{{\cal L}_+})^{\mu}_{\;\;0}+(\Lambda_{{\cal L}_+})^{\mu}_{\;\;3}\ ,\\ 
    \alpha_{_{{\cal L}_+}}^{-1} &:~ {\cal L}_+ \rightarrow \mathrm{SO}(1,3)/\mathrm{ISO}(2)\ , \quad   \hspace{1.9cm}   y^\mu \mapsto [\Lambda_{{\cal L}_+}]\ ,
\end{aligned}
\end{equation}
where, again, $[\Lambda_{{\cal L}_+}]=\{\Lambda\in\mathrm{SO}(1,3):\Lambda\sim\Lambda_{{\cal L}_+}\}$ is the equivalence class (under right ISO(2)-multiplication) of the representative $\Lambda_{{\cal L}_+}$. Here we cannot directly write an expression for $\Lambda_{{\cal L}_+}$ similar to \eqref{matrixT} and \eqref{matrixS}, but, for each equivalence class, we can still write a simple representative parametrized by the $y$-coordinates, namely
\begin{equation}\label{matrixL}
    \Lambda_{{\cal L}_+} \= \begin{pmatrix} \frac{y^0}{2}{+}\frac{1}{y^0{+}y^3} & \frac{y^1}{y^0{+}y^3} & \frac{y^2}{y^0{+}y^3} & \frac{y^0}{2}{-}\frac{1}{y^0{+}y^3} \\[4pt] \frac{y^1}{2} & 1 & 0 & \frac{y^1}{2} \\[4pt] \frac{y^2}{2} & 0 & 1 & \frac{y^2}{2} \\[4pt] \frac{y^3}{2}{-}\frac{1}{y^0{+}y^3} & -\frac{y^1}{y^0{+}y^3} & -\frac{y^2}{y^0{+}y^3} & \frac{y^3}{2}{+}\frac{1}{y^0{+}y^3} \\ \end{pmatrix}\ .
\end{equation}

The Maurer--Cartan one-forms can be obtained from $\Lambda_{{\cal L}_+}^{-1}\text{d}\Lambda_{{\cal L}_+} {=}\, e^a I_a$ (no $I_i$ terms, since $\frak{m}$ here is actually a subalgebra of $\frak{g}$) and expressed in terms of the $y$-coordinates, but they read better in terms of spherical spatial coordinates, with
\begin{equation}
    y^1 = r\,\sin{\theta} \cos{\varphi}\ ,\quad y^2 = r\,\sin{\theta} \sin{\varphi}\ ,\quad y^3 = r\,\cos{\theta}\ ,\quad \text{and} \quad y^0 = r\ ,
\end{equation}
then one obtains
\begin{equation}
    e^1 = \tfrac{r}{2}\left( \cos{\varphi}\,\dd\theta{-}\sin{\theta}\sin{\varphi}\,\dd\varphi \right)\ ,\quad e^2 = \tfrac{r}{2}\left( \sin{\varphi}\,\dd\theta{+}\sin{\theta}\cos{\varphi}\,\dd\varphi \right)\ ,\!\!\und e^3 = \tfrac{1}{r}\,\dd r {-} \tan\sfrac{\theta}{2}\,\dd\theta\ .
\end{equation}
They provide a local orthonormal frame on ${\cal L}_+$, such that
\begin{equation}
    \dd s^2_{\R^{1,3}} \big|_{{\cal L}_+} \= 4\, e^1\otimes e^1 + 4\, e^2\otimes e^2\ .
\end{equation}
The metric in this case is degenerate (the $e^3\otimes e^3$ term vanishes), as expected for the lightcone.

One can also explicitly write the action of the generator $K_3$ on the base vector,
\begin{equation}
\ep^{u K_3}\cdot \begin{pmatrix} 1 \\ 0 \\ 0 \\ 1 \end{pmatrix} \= \begin{pmatrix} \text{e}^{u} \\ 0 \\ 0 \\ \text{e}^{u} \end{pmatrix},
\end{equation}
to see that $K_3$ generates dilatations on the lightcone. This means that the orbit of the coset on a base vector $(\ep^u,0,0,\ep^u)^\top$ coincides with the orbit of $(1,0,0,1)^\top$ for any $u\in\R$; this stands in contrast with the previous cases.  Actually, the map $\alpha_{{\cal L}_+}$ in \eqref{maplightcone} is onto, so any base vector on the future lightcone generates the whole future lightcone. Moreover, there is no foliation of the lightcone here, which means that $\mathrm{SO}(1,3)$-invariant Yang--Mills fields in the lightcone have no dynamics. The gauge field in this case is necessarily pure gauge, with $\Fcal=0$.

\end{document}